\documentclass{article}

\newcommand{\bt}{\begin{tabular}{c}}
\newcommand{\et}{\end{tabular}}

\newcommand{\eb}{\ee\be } 
 
\newcommand{\bmat}{\lt ( \begin{array} }
\newcommand{\emat}{  \end{array} \rt )}

\newcommand{\oq}{{\ov \q}}

\newcommand{\oy}{{\ov \y}}

\newcommand{\oF}{{\ov F}}
\newcommand{\A}{{\ov A}}

\renewcommand{\a}{\alpha}	
\renewcommand{\b}{\beta}
\newcommand{\g}{\gamma}
\renewcommand{\d}{\delta}
\newcommand{\e}{\epsilon}

\newcommand{\z}{\zeta}

\newcommand{\q}{\theta}

\newcommand{\m}{\mu}

\newcommand{\y}{\psi}
\newcommand{\w}{\omega}

\newcommand{\D}{\Delta}
	
\renewcommand{\L}{\Lambda}

\newcommand{\la}{\label}
\newcommand{\ci}{\cite}

\newcommand{\ds}{\documentstyle}	
\newcommand{\fr}{\frac}

\newcommand{\pa}{\partial}
\newcommand{\ov}{\overline}
\newcommand{\be}{\begin{equation}}
\newcommand{\ee}{\end{equation}}
\newcommand{\ba}{\begin{array}} 
\newcommand{\ea}{\end{array}}
\newcommand{\bea}{\begin{eqnarray}}
\newcommand{\eea}{\end{eqnarray}}
\newcommand{\ra}{\rightarrow}
\newcommand{\Ra}{\Rightarrow}
\newcommand{\lra}{\longrightarrow}

\newcommand{\lt}{\left}
\newcommand{\rt}{\right}

\newcommand{\ben}{\begin{enumerate}}
\newcommand{\een}{\end{enumerate}}
\newcommand{\bitem}{\begin{itemize}}
\newcommand{\eitem}{\end{itemize}}
\newcommand{\articlenumber}{}

\newcommand{\articletitle}{ Supersymmetry Breaks when Gauge Symmetry Breaks:\\ Cybersusy I }

\begin{document}
\makeatletter	   
\renewcommand{\ps@plain}{%
\renewcommand{\@oddhead}{{\articlenumber  \hspace{1cm} }\hspace{1cm}  \hfil\textrm{\thepage}} 
\renewcommand{\@evenhead}{\@oddhead}
\renewcommand{\@oddfoot}{\textrm{\articlenumber \hspace{1cm}
  }\hspace{1cm} \hfil\textrm{\thepage}}
\renewcommand{\@evenfoot}{\@oddfoot}}
\makeatother    
\title{ \articletitle\\ \articlenumber}
\author{ J. A.  Dixon\footnote{jadix@telus.net}\\ Dixon Law Firm\footnote{Fax: (403) 266-1487} \\1020 Canadian Centre\\
833 - 4th Ave. S. W. \\ Calgary, Alberta \\ Canada T2P 3T5 }
\maketitle
\pagestyle{plain}
\Large

\abstract{\large
This paper summarizes  a new approach to supersymmetry breaking  in the supersymmetric standard model (SSM). The approach arises from some remarkable features of the BRS cohomology for composite operators in the SSM, and the behaviour of those operators when gauge symmetry is spontaneously broken. A new realization of supersymmetry arises for these operators.  This realization is equivalent to the generation of supersymmetry anomalies, though they are not present in the usual sense.  

The consequences are worked out in detail for the electron and  neutrino flavour triplets, by using the appropriate effective action to analyze the new anomalous realization of supersymmetry. 

This effective action generates a mass spectrum for leptons that 
is consistent with present experimental data. 
 There is no  vacuum energy problem, and  no annoying  mass sum rules are   present.

\Large

\section{Introduction}

Thirty-four years have passed since the discovery of the theory of four dimensional supersymmetry in \ci{WZ}. The excitement that it initially created gave rise to many interesting and useful early papers, some of which are collected in  \ci{ferrara}.     Supersymmetry is still appealing to many physicists, as is evident from the proceedings of the recent conferences on SUSY \ci{SUSY06} \ci{SUSY07} \ci{SUSY08}. 

With the advent of the LHC, we may soon have some experimental evidence for supersymmetric partners to the known particles.
But how can we recognize any new particles as supersymmetric partners of known particles? There is a gap between conserved supersymmetry and the real world.  This is the theoretical problem of supersymmetry breaking, and many physicists feel that it has not yet been resolved in a satisfactory way.

This  paper  describes a new approach to supersymmetry breaking.  It appears that this approach resolves the problem of comparing theory with experiment, without introducing other problems.  A first sketch of the phenomenology  can be seen below in section 
\ref{qfggqegerujujkuy}.  That section deals with the
 electrons, muons and tau leptons, and their superpartners, for the special case where all the mixing matrices are diagonal. The more general case is worked out in \ci{cybersusyIV}.  
The neutrinos behave in the same way, but with different matrices, reflecting their smaller masses.

The new approach is based on some remarkable features of the BRS cohomology of supersymmetry applied to composite operators. It will be seen that supersymmetry is broken by those operators.  The breaking  happens when the vacuum expectation value (VEV) is turned on. So supersymmetry and gauge symmetry break together. 
The breaking is closely related  to supersymmetry anomalies that resemble the the well-known gauge anomalies, but there are also big differences from those anomalies. 

Essentially, what happens is that certain important composite operators develop a new realization of supersymmetry. This new realization is very specific to the  standard supersymmetric model (SSM).  The new realization   incorporates   effective supersymmetry anomalies proportional to the VEV. We call this new anomalous algebra the `cybersusy algebra'. Experimental predictions from this new algebra can be made using the technique of effective actions. A quick sketch of the historical development of this subject is attempted  in section 
\ref{history} below, and the origin of the name `cybersusy' is discussed  in section \ref{fsdfsdfdfsdf}. 
 The calculations that stand behind the results of this introductory paper  can be found in three papers that follow this introductory paper.  They are 
\ci{cybersusyII}
\ci{cybersusyIII}
\ci{cybersusyIV}.

\section{Notation for the Supersymmetric Standard Model}

\la{notationsub}

Here is the superspace potential for the Supersymmetric Standard Model (SSM):
\be
P_{{\rm SP}}   =
g \e_{ij} H^i K^j J
- g_{\rm J} m^2 J
+
p_{p q} \e_{ij} L^{p i} H^j P^{ q} 
\eb
+
r_{p q} \e_{ij} L^{p i} K^j R^{ q}
+
t_{p  q} \e_{ij} Q^{c p i} K^j T_c^{ q}
+
b_{p  q} \e_{ij} Q^{c p i} H^j B_c^{ q}
\la{fqwefweef1212}
\ee

The massless SSM arises when we set $g_{\rm J} m^2=0$ in the Superpotential (\ref{fqwefweef1212}) above.   We are forced to introduce neutrino masses through the right handed neutrino superfield $R^p$ to implement the present supersymmetry breaking mechanism for neutrinos.

All of the above fields in  $P_{\rm SP} $ are the scalar parts of chiral superfields.    When the scalars are replaced by their superfields and $P_{\rm SP}$ is put into the SSM action,  the $g_J$ term gives rise to the VEV which spontaneously breaks the gauge symmetry $SU(3) \times SU(2) \times U(1)$
 down to $SU(2) \times  U(1)$:

\be
g v^2 = g_J
\ee
\be
<H^1> = 
<K_1> = mv
\ee

 We use ${\widehat E}^q$   to denote a superfield whose $\q,\oq$ independent part is the scalar $E^q$  . For example $L^{iq}$ is the scalar part of the left leptonic SU(2) doublet chiral scalar superfield 
${\hat L}^{iq}$.  The spinor component is labelled $\y_{L \a}^{iq }$
and the auxiliary is $F_{L }^{iq}$.  The complex conjugate is 
${\widehat {\ov L}}_{iq}$.  The scalar component  of ${\widehat {\ov L}}_{iq}$ is labelled ${\ov L}_{iq}$, the spinor component  of ${\widehat {\ov L}}_{iq}$ is labelled $\oy_{L i q \dot \a}$
and the auxiliary is $\ov F_{L i q}$.  A chiral scalar superfield ${\widehat L}^{iq}$ satisfies the chiral constraint 
${\ov D}_{\dot \a} {\widehat L}^{iq}=0$, and its complex conjugate satisfies the antichiral constraint ${ D}_{ \a} {\widehat {\ov L}}_{iq}=0$.   Our notation is based on  that in \ci{superspace}, though there are differences. A summary of our notation is in the appendix to \ci{cybersusyII}.

 After spontaneous breaking of gauge symmetry, the two $SU(2)$ doublets will give rise to the following superfields:
\be
Q^{c p i} \ra 
\lt (
\begin{array}{c}
U^{c p} 
\\
D^{c p} 
\\
\end{array}
\rt )
; L^{p i} \ra
\lt (
\begin{array}{c}
N^{p} 
\\
E^{p } 
\\
\end{array}
\rt )
\la{geghhgehhephtp}
\ee
Here is an aid to remembering the notation: 
\ben
\item
 ${\widehat Q}^{c pi}$ is the Quark doublet superfield, and  
\item
 ${\widehat L}^{pi}$ is the Lepton doublet superfield, and  
\item
 ${\widehat U}^{c p} $ and ${\widehat T}_c^{ p}$ are the left and right `Up-Top' squark superfields, and  
\item
${\widehat D}^{c p} $ and ${\widehat B}_c^{ p}$ are the left and right `Down-Bottom' squark superfields. 
\item
${\widehat N}^{p} $ and ${\widehat R}^{ p}$ are the left and right `Neutrino-Right' leptonic superfields,   and
\item
${\widehat E}^{p } $ and ${\widehat P}^{ p}$ are the left and right `Electron-Positron' leptonic superfields,   
\item
${\widehat H}^i$, ${\widehat J}$ and ${\widehat K}^i$ are the Higgs-type superfields,   
\item
The complex  matrices $p_{pq}$,  $r_{pq}$,  $t_{pq}$,  $b_{pq}$ are matched to, and named after, the right handed SU(2) singlet superfields ${\widehat P}^{q}$, ${\widehat R}^{q}$, ${\widehat T}^{cq}$,
 ${\widehat B}^{cq}$,  to form interaction terms, which also become mass terms when the gauge symmetry is spontaneously broken.
\item
 Note that the second index of $p_{pq}$ is contracted with the right superfield ${\widehat P}^{q}$, etc.
\een
Here is a table summarizing the various quantum numbers:
\be
\begin{tabular}{|c|c|c
|c|c|c
|c|c|c|}
\hline
\multicolumn{8}
{|c|}{Table of the Chiral Superfields in the SSM
}
\\
\hline
\multicolumn{8}
{|c|}{ \bf Superstandard Model, Left ${\cal L}$
Fields}
\\
\hline
{\rm Field} & Y 
& {\rm SU(3)} 
& {\rm SU(2)} 
& {\rm F} 
& {\rm B} 
& {\rm L} 
& {\rm D} 
\\
\hline
$ L^{pi} $& -1 
& 1 & 2 
& 3
& 0
& 1
& 1
\\
\hline
$ Q^{cpi} $ & $\fr{1}{3}$ 
& 
3 &
2 &
3 &
 $\fr{1}{3}$
& 0
& 1
\\\hline
$J$
& 0 
& 1
& 1
& 1
& 0
& 0
& 1
\\
\hline
\multicolumn{8}
{|c|}{ \bf Superstandard Model, Right 
${\cal R}$
Fields}
\\
\hline
$P^{ p}$ & 2 
& 1
& 1
& 3
& 0
& -1
& 1
\\
\hline
$R^{ p}$ & 0 
& 1
& 1
& 3
& 0
& -1
& 1
\\

\hline
$T_c^{ p}$ & $-\fr{4}{3}$ 
& ${\ov 3}$ &
1 &
3 &
 $-\fr{1}{3}$
& 0
& 1
\\
\hline
$B_c^{ p}$ & $\fr{2}{3}$ 
& ${\ov 3}$ 
&
1 &
3 &
 $- \fr{1}{3}$
& 0
& 1
\\
\hline
$H^i$ 
& -1 
& 1
& 2
& 1
& 0
& 0
& 1
\\
\hline
$K^i$ 
& 1 
& 1
& 2
& 1
& 0
& 0
& 1
\\
\hline
\end{tabular}
\\
\la{qefewfqwefwrrr}
\ee
\vspace{.2cm}
In the above, Y is weak hypercharge, F stands for the number of families for each superfield, B is baryon number, L is lepton number and D stands for mass dimension.

\section{Some Special Composite Operators in the SSM}

Consider the following composite operators\footnote{In these expressions $q$ is a flavour index, $i$ on $K^i$ is an SU(2) isospin index and $\dot \a$ is a Weyl dotted spinor index.} in the supersymmetric standard model (SSM):
\be
\w_{P q\dot \a} = g \oy_{L i q \dot \a} K^i  + p_{qp}\oy_{J \dot \a} 
 P^{  p} +  \cdots 
\la{fwewfewef1}
\ee
\be
\w_{E p\dot \a} = g \oy_{P  p \dot \a} K^i  H_i -p_{qp}\oy_{J \dot \a} 
 L^{i  q} H_i +  \cdots 
\la{fwewfewef2}
\ee
 These   composite operators are very special, because:
\ben
\item
When the scalar fields $H^i$ and $K^i$ get their vacuum expectation values (VEVs), the leading terms in the above operators are the elementary spinor fields for the antielectron ($\oy_{E q \dot \a}$)  and the antipositron ($\oy_{P q \dot \a}$) times coupling constants times powers of the mass parameter $m$.  Hence these operators clearly are physical, at least after gauge symmetry gets broken.
\item
The BRS cohomology of the massless SSM \ci{cybersusyIII} tells us how to fill in the dots $+ \cdots $ in the operators (\ref{fwewfewef1}) and 
(\ref{fwewfewef2}) so as to make these into some very special composite operators that behave under supersymmetry like chiral dotted spinor superfields  ${\widehat \w}_{E p\dot \a}$ and 
${\widehat \w}_{P q\dot \a}$. A chiral dotted spinor superfield ${\widehat \w}_{E q\dot \b}$ satisfies the chiral constraint 
${\ov D}_{\dot \a} {\widehat \w}_{E q\dot \b}=0$. The terms  (\ref{fwewfewef1}) and 
(\ref{fwewfewef2}) are just the $\q,\ov \q$ independent parts of those chiral dotted spinor superfields, in accord with our usual notation in section \ref{notationsub}.
\item
When the VEVs appear in the SSM, the chiral dotted  composite superfields  that result from  (\ref{fwewfewef1}) and 
(\ref{fwewfewef2})  cease to behave like superfields. Under the action of the BRS operator $\d$ for supersymmetry, they have the usual supersymmetry variations plus the following:
\be
\d {\widehat \w}_{P p\dot \a}= g m^2 v^2 p_{pq}  
{\widehat P}^{q}
\ov C_{\dot \a}
\la{ewrwerwerP}
\ee
\be
\d {\widehat \w}_{E p\dot \a}= g m^2 v^2 p_{qp}\lt (
m v {\widehat E}^{q}
+ \cdots\rt )\ov C_{\dot \a}
\la{ewrwerwerE}
\ee
where $ {\widehat E}^{q}$ is the chiral scalar superfield for the electron,
and $\lt (
m v {\widehat E}^{q}
+ \cdots\rt )$ is a composite superfield, and  ${\widehat P}^{q}$ is the chiral scalar superfield for the positron
\item
Note that in the spontaneously broken theory, $ {\widehat \w}_{P p\dot \a}$ 
and ${\widehat P}^{q}$ both have the charge of the positron, and 
$ {\widehat \w}_{E p\dot \a}$ 
and ${\widehat E}^{q}$ both have the charge of the electron. \item
Perhaps it is best to think of the multiplet $ {\widehat \w}_{P p\dot \a}$  as a supersymmetric set of bound states of the antielectron multiplet with the scalars $K^i$, following equation (\ref{fwewfewef1}), with a similar notion for $ {\widehat \w}_{E p\dot \a}$, but this is not at all clear, nor is it necessary to decide this issue for present purposes. Anyway these `bound states' get mixed up with the elementary fields.  
\item
However, as will be seen later, this concept of supersymmetric bound states, driving supersymmetry breaking, becomes much more plausible and compelling for baryons.  Indeed, it becomes irresistible for baryons.  But it is not yet known whether cybersusy makes sense for baryons, because the necessary calculations are long and have not yet been done.

\een
\section{An attempt to formulate a more general BRS version of the philosophy of effective fields and effective actions}
\la{fqwfwefwfwef}
There is a new and important realization of the supersymmetry algebra embodied in the transformations (\ref{ewrwerwerP}) and 
(\ref{ewrwerwerE}).  In order to explore that realization we use the technique of effective actions.  

Let us repeat here the following useful remarks from the second volume of Weinberg's treatise on quantum field theory
\ci{weinberg2effectiveaction},
which 
summarizes the philosophy of using  effective actions in a way that seems relevant to what we are going to do:  

{\em Some years after ... there emerged a different justification for the effective field theory technique...
It is based on the realization (not yet formally embodied in a theorem) that when we calculate a physical amplitude from Feynman diagrams using the most general Lagrangian that involves the relevant degrees of freedom and satisfies the assumed symmetries of the theory, we are simply constructing the most general amplitude that is consistent with general principles of relativity, quantum mechanics and the assumed symmetries. }

Here we are going to start with the philosophical remark above and then take the following steps, inspired by the above remarks, but adapted to what happens with the current situation:
\ben
\item
Define elementary `effective' fields, with canonical dimensions, which  correspond to the composite fields.  
\item
The canonical dimensions of the effective fields are to be determined so that the kinetic part of the action, which is to be constructed in  item
 \ref{rqgtggreqg} below,  contains those fields  normalized so that the pure kinetic part does not contain factors of mass. 

\item
Write down a BRS operator $\d$ for the elementary `effective' fields which has the same   symmetry transformations and the same algebra as the composite fields. 
\item
Check that the BRS algebra closes:
\be
\d^2 = 0.
\ee
\item
Write down the most general local Lagrangian field theory which contains those effective fields and which is invariant under that $\d$, and also examine related field theories, made of the same fields, that are invariant under related nilpotent operators $\d$.  
\la{rqgtggreqg}
\item
Use those local Lagrangian field theories to derive information  about the symmetry characterized by $\d$, its properties, and, when appropriate, the breaking of that symmetry.
\een

\section{ Cybersusy Algebra for the Electrons}
\la{cyberalgebrasection}

We will call the present approach cybersusy, for reasons explained in section \ref{fsdfsdfdfsdf}.
Let us try to carry out these steps in the above order, and see what happens:
\ben
\item
First we assign new elementary effective chiral superfields to the old composite superfields, as  follows. We could use the same names for the first fields
 (\ref{fqfwfewerL}) and (\ref{fqfwfewerR}), but that would confuse composite with elementary fields, so we will use new names. We will use the subscripts L(eft) and R(ight)  to denote the effective 
superfields\footnote{ The  same effective superfields reappear in other contexts--the neutrino is identical except that $p_{pq} \ra r_{pq}$,  and  the\\ hadrons will have this algebra as part of a larger algebra.}:
\be
{\widehat \w}_{E p\dot \a} \Ra 
{\widehat \w}_{L p\dot \a}  
\la{fqfwfewerL}
\ee\be
{\widehat \w}_{P p\dot \a} \Ra 
{\widehat \w}_{R p\dot \a}  
\la{fqfwfewerR}\ee
\be
g m^2 v^2 {\widehat E}^p  \Ra 
{\widehat A}_{L}^{ p }  
\la{qgregegergL}\ee
\be
g m^2 v^2 \lt (
m v {\widehat P}^{q}
+ \cdots\rt ) \Ra 
{\widehat A}_{R}^{ p }  
\la{qgregegergR}\ee
\item

The masses in (\ref{qgregegergL}) and (\ref{qgregegergR}) have been absorbed.  All the fields get new dimensions, to be consistent with the canonical dimensions for these new effective chiral superfields, which are as follows:
\be
{\widehat \w}_{L p\dot \a} 
,{\widehat \w}_{L p\dot \a} 
\;{\rm have \; Dimension} = \fr{1}{2}
\ee
\be
{\widehat A}_{L}^{p} 
,{\widehat A}_{R}^{p} 
\;{\rm have \; Dimension} = 1
\ee
and we note that
\be
\ov C_{\dot \a} 
, C_{ \a}  
\;{\rm have \; Dimension} = -\fr{1}{2}
\ee
\be
m, \pa_{\a \dot \b}
\;{\rm have \; Dimension} = 1
\ee
These dimensions are canonical for the action to be written down below in section \ref{qegregregergre}.

\item
Next we write down the abstract algebra as realized on these new fields.  We can use chiral superfields to do this.  The BRS operator divides into two parts
\be
\d = \d_{\rm SS} + 
 \d_{\rm GSB} 
\la{qggrqegqerg1}
\ee
\ben
\item
 The part $\d_{\rm SS} $ incorporates the usual supersymmetry 
transformations on the superfields:
\be
 \d_{\rm SS}  {\widehat \w}_{L p\dot \a}=\lt \{
 C_{\a} Q^{\a}     + \ov C_{\dot \a} \ov Q^{\dot \a}
\rt \}{\widehat \w}_{L p\dot \a}
\ee
\be
 \d_{\rm SS}  {\widehat \w}_{R p\dot \a}=\lt \{
 C_{\a} Q^{\a}     + \ov C_{\dot \a} \ov Q^{\dot \a}
\rt \}{\widehat \w}_{R p\dot \a}
\ee
\be
 \d_{\rm SS}  {\widehat A}_{L }^{p}=\lt \{
 C_{\a} Q^{\a}     + \ov C_{\dot \a} \ov Q^{\dot \a}
\rt \}{\widehat A}_{L }^{p}
\ee
\be
 \d_{\rm SS}  {\widehat A}_{R }^{p}=\lt \{
 C_{\a} Q^{\a}     + \ov C_{\dot \a} \ov Q^{\dot \a}
\rt \}{\widehat A}_{R }^{p}
\ee
\item
\la{deltaGSB}
The part $\d_{\rm GSB} $ (the GSB stands for `gauge symmetry breaking') incorporates the peculiar modification that arises in equations (\ref{ewrwerwerP}) and 
(\ref{ewrwerwerE}) from gauge symmetry breaking when the VEV is non-zero ($p_{qp}$ is dimensionless) :

\be
 \d_{\rm GSB}  {\widehat \w}_{L p\dot \a}=  p_{qp} {\widehat A}_{L }^{q}
\ov C_{\dot \a}
\la{deltaGSBL}
\ee
\be
 \d_{\rm GSB}  {\widehat \w}_{R p\dot \a}=   p_{pq}  
{\widehat A}_{R }^{q}
\ov C_{\dot \a}
\la{deltaGSBR}
\ee
\be
 \d_{\rm GSB}  {\widehat A}_{L }^{p}=  0
\ee
\be
 \d_{\rm GSB}  {\widehat A}_{R }^{p}=  0
\la{qggrqegqerg2}
\ee
\een
\item
Note that we can divide the operator $\d$ into independent left and right parts which act on the left and right superfields:
\be
\d 
=\d_{\rm L}
+ \d_{\rm R}
= \d_{\rm SS} + 
 \d_{\rm GSB} \ee
where
\be
\d_{\rm L} = \d_{\rm SSL} + 
 \d_{\rm GSBL} 
\ee
\be
\d_{\rm R} = \d_{\rm SSR} + 
 \d_{\rm GSBR} 
\ee
and
\be
\d_{\rm SS} = \d_{\rm SSL} + 
 \d_{\rm SSR} 
\ee
\be
\d_{\rm GSB} = \d_{\rm GSBL} + 
 \d_{\rm GSBR} 
\ee
The left and right parts of $\d$ are disconnected and we can further divide them in various ways.
It is easy to verify that many of these suboperators are also nilpotent. For example:
\be
\d^2  =\d_{\rm SS}^2=\d_{\rm GSB}^2=\d_{\rm SSL}^2=\d_{\rm GSBL}^2=\d_{\rm L}^2 =\d_{\rm R}^2= 0
\ee
\item
In \ci{cybersusyIV}, we will return to this algebra and write it out in components.  That is desirable, since as will be seen below, we must use components to write out the action which is invariant under the cybersusy algebra (see 
(\ref{formKCL1}) below for example), and components are essential when working out the masses, since supersymmetry is entirely gone at that point.  
\een

\section{ Cybersusy Action  for the pure supersymmetry operator $\d = \d_{\rm SS}$ when the VEV is zero}

\la{qegregregergre}
Now we write down the general local Lagrangian field theory that uses these fields and that embodies this symmetry and the various subsymmetries. We shall only be interested here in the simplest part of that, namely the free kinetic part.  

First we will write down the action for the special case where 
the transformations are ordinary supersymmetry, so that $\d_{GSB}=0$. We  use superspace notation to compactify this and then we write the actions out in components:
We want
\be
 \d_{\rm SS} {\cal A}_{\rm Susy} =0
\ee
The canonical general action for this is:
\[
{\cal A}_{\rm Susy} ={\cal A}_{\rm  ScalarL}
+{\cal A}_{\rm  ScalarR}
+{\cal A}_{\rm Scalar\;  Mass} 
\]
\be
+
{\cal A}_{\rm  DotspinorL}
+
{\cal A}_{\rm  DotspinorR}
+
{\cal A}_{\rm Dotspinor\;  Mass} 
\ee
where the scalar action consists of three parts:
\[
{\cal A}_{\rm  ScalarL}
=\fr{1}{4}\int d^4 x \; d^4 \q
{\widehat A}_{L }^{p} 
{\widehat {\ov A}}_{L p} \]
\be
=
 \int d^4 x \;
\lt (
- A^{ p  }_{{L} }
\D \A^{ }_{{L} p  }
-
 \y^{    \a p}_{{L} }
\pa_{\a \dot \b}
 \oy^{ \dot \b}_{{L} p   }
+
 F^{  p }_{{L} }
 \oF^{ }_{{L} p  }
\rt )
\la{ActionScalarL}
\ee

\[
{\cal A}_{\rm  ScalarR}=
\fr{1}{4}
\int d^4 x \; d^4 \q
{\widehat A}_{R }^{p} 
{\widehat {\ov A}}_{R p} 
\]\be
=
\int d^4 x \;
\lt (
- A^{ { p }   }_{{R} }
\D \A^{ }_{{R} { p }   }
-
 \y^{    \a { p } }_{{R} }
\pa_{\a \dot \b}
 \oy^{ \dot \b}_{{R} { p }    }
+
 F^{  { p }  }_{{R} }
 \oF^{ }_{{R} { p }   }
\rt )
\la{ActionScalarR}
\ee

\[
{\cal A}_{\rm Scalar\;  Mass}  =
\fr{1}{4}\int d^4 x \; d^2 \q
  g_{pq} m 
{\widehat A}_{L}^{p} 
{\widehat A}_{R}^{q} 
\]
\[
+
\fr{1}{4}\int d^4 x \; d^2 \ov \q
\ov g^{pq} m {\widehat {\ov A}}_{L  p} 
{\widehat {\ov A}}_{R q } 
\]
\[
=\int d^4 x \;
 m g_{p q}
\lt (
- \y^{ \a  p}_{{L} }
 \y^{ q}_{{R} \a}
+
 A^{  p }_{{L} }
 F^{ q}_{{R} }
+
 F^{ p }_{{L} }
 A^{ q}_{{R}}
\rt )
\]
\be+
\int d^4 x \;
 m \ov g^{p  q}
\lt (
- \oy^{\dot \a  }_{{L}p }
 \oy^{}_{{R} \dot \a  q}
+
 \A^{   }_{{L} p}
 \ov F^{}_{{R}  q}
+
 \ov F^{  }_{{L} p}
 \A^{}_{{R}  q}
\rt )
\la{ActionScalarMass}
\ee
and the dotspinor action also consists of three parts:
\[
{\cal A}_{\rm  DotspinorL}
=-\fr{1}{4}\int d^4 x \; d^4 \q
{\widehat {\ov \w}}_{L  \a }^{p} 
\pa^{\a \dot \b}
{\widehat \w}_{L p\dot \b} 
\]
\be
=
\int d^4 x \;
\lt (
 \ov \w^{ \a  p }_{{L}  }
\pa_{\a \dot \b}
\D 
\w^{  \dot \b }_{{L}   
 p }
-
 \ov W^{ \a \dot \g  p }_{{L}  }
\pa_{\a \dot \b} 
\pa_{\d \dot \g}
W^{  \d \dot \b}_{{L}   
 p  }
-
\ov \L^{ p  \a}_{{L}  }
\pa_{\a \dot \b}
 \L^{  \dot \b}_{{L}   
 p  }
\rt )
\la{ActionSpinorL}
\ee

\[
{\cal A}_{\rm  DotspinorR}
=\fr{1}{4}
\int d^4 x \; d^4 \q
{\widehat {\ov \w}}_{R  \a }^{p} 
\pa^{\a \dot \b}
{\widehat \w}_{R p\dot \b} 
\]
\be=
\int d^4 x \;
\lt (
 \ov \w^{ \a {{p}}}_{{R}  }
\pa_{\a \dot \b}
\D 
\w^{  \dot \b }_{{R}   
{{p}}}
-
 \ov W^{ \a \dot \g {{p}}}_{{R}  }
\pa_{\a \dot \b} 
\pa_{\d \dot \g}
W^{  \d \dot \b}_{{R}   
{{p}} }
-
\ov \L^{{{p}} \a}_{{R}  }
\pa_{\a \dot \b}
 \L^{  \dot \b}_{{R}   
{{p}} }
\rt )
\la{ActionSpinorR}
\ee

\[
{\cal A}_{\rm Dotspinor\;  Mass}  =
-\fr{1}{4}\int d^4 x \; d^2 \ov \q
\ov d_{pq} m^2 {\widehat {\ov \w}}_{L  \a }^{p} 
{\widehat {\ov \w}}_{R  }^{ \a q} 
\]\[
-
\fr{1}{4}\int d^4 x \; d^2 \q
-  d^{pq} m^2 
{\widehat \w}_{L p\dot \b} 
{\widehat \w}_{R q}^{\dot \b} 
\]
\[
=
\int d^4 x \;
  m^2 d^{ p  q}  
\lt (
- \w^{\dot \a  }_{{L}  p }
 \L^{}_{{R}\dot \a q}
-
 \L^{\dot \a  }_{{L}  p }
 \w^{}_{{R} \dot \a q}
-
 W^{\a \dot \a  }_{{L}  p }
 W^{}_{{R} \a \dot \a q}
\rt )
\]
\be
+
\int d^4 x \;
    m^2 \ov d_{ p  q}  
\lt (
 - \ov \w^{\a  p  }_{{L} }
 \ov \L^{ \ q}_{{R}\a }
-
 \ov \L^{ p  \a  }_{{L}}
 \ov \w^{   q}_{{R}  \a }
-
 \ov W^{  p \a \dot \g  }_{{L} }
 \ov W^{  q}_{{R} \a \dot \g }
\rt )
\la{ActionSpinorMass}
\ee

These kinetic and mass terms for the scalar superfields ${\widehat A}_{L}^{p} $ and $
{\widehat A}_{R}^{q} $ are familiar, but, as far as this author  knows\footnote{Chiral dotted superfields  are also discussed in   \ci{superspace}, where a constraint is imposed to  extract a vector represention.}, the 
kinetic and mass terms for the chiral dotted spinor superfields
 ${\widehat \w}_{L p\dot \b} $ and $
{\widehat \w}_{R q}^{\dot \b} 
$ are new in the above.   
Note that the left and right actions above are invariant under left and right parts of $\d_{SS} $ separately. For example 
\be
\d_{SSL} {\cal A}_{\rm  ScalarL}
=\d_{SSL} {\cal A}_{\rm  DotspinorL}
=0
\ee However the mass terms mix left and right, and they are invariant only under the entire operator $\d_{SS} $ with both left and right parts.

\section{Equations of motion before gauge symmetry breaking}
\subsection{Equations of motion of Scalar Superfields}
First let us have a look at the equations of motion for the chiral scalar field from the above action.  
The action for 
 the scalar superfields is:
\be
{\cal A}_{\rm   Scalar}=
{\cal A}_{\rm  ScalarL}
+{\cal A}_{\rm  ScalarR}
+{\cal A}_{\rm Scalar\; Mass }
\la{ActionScalar}
\ee
where these are defined in (\ref{ActionScalarL}), 
 (\ref{ActionScalarR}) and (\ref{ActionScalarMass}).
 The equations of motion follow easily from functional derivations.
One writes down all the equations like 
\be
\fr{\d {\cal A}_{\rm  Scalar }}{\d A^{ p  }_{L }}=0
=- \D \A^{ }_{{L} p  }
+
 m g_{r p}
 F^{ r }_{{R} }
\ee
\be
\fr{\d {\cal A}_{\rm  Scalar }}{\d  \oF^{ }_{{R} { p }   }}=0
=
 F^{  { p }  }_{{R} }
+
 m \ov g^{s  p}
 \A^{   }_{{L} s}
\ee
  and then one needs to perform some operations to combine the equations and one gets equations like
\be
 \D \A^{ }_{{L} p  }
+
 m^2 g_{r p}
  \ov g^{s  r}
 \A^{   }_{{L} s}=0
\la{qerwfgegqerger}
\ee
Since the matrix $G_{p}^{\;\;s} = g_{r p}
  \ov g^{s  r}$ is a positive definite hermitian matrix, it can be diagonalized and this yields three equations of the form
\be
\lt ( \D 
+
 m^2 G
 \rt )
 \A^{   }_{{L}}0
\ee
for the three flavours, with $G$ different for each flavour.

\subsection{Equations of motion of Dotspinor Superfields}

Next, let us have a look at the equations of motion for the chiral dotspinor field from the above action.  
The action for 
 the dotspinor superfields is:
\be
{\cal A}_{{\rm   Dotspinor} }=
{\cal A}_{\rm  DotspinorL}
+
{\cal A}_{\rm  DotspinorR}
+{\cal A}_{\rm Dotspinor\;  Mass} 
\ee
where these are defined in (\ref{ActionSpinorL}), (\ref{ActionSpinorR}) and (\ref{ActionSpinorMass}).

Again here the equations of motion follow easily from functional derivations.  But there are no auxiliaries to 
 eliminate.  And there is an extra factor of $\D$ in many places.
For example:
\be
\fr{\d {\cal A}_{\rm Dotspinor }}{\d  \ov W^{ \a \dot \g  p }_{{L}  }
}=0
=
-\pa_{\a \dot \b} 
\pa_{\d \dot \g}
W^{  \d \dot \b}_{{L}   
 p  }
-
    m^2 \ov d_{ p  q}  
 \ov W^{  q}_{{R} \a \dot \g }
\la{fweweffwef1}
\ee
and
\be
\fr{\d {\cal A}_{\rm Dotspinor }}{\d  W^{  }_{{R \z \dot \e}   
{{q}} }}=0=
-\pa^{\a \dot \e} 
\pa^{\z \dot \g}
 \ov W^{ {{q}}}_{{R} \a \dot \g  }
-
  m^2 d^{ s  q}  
 W^{\z \dot \e}_{{L}  s   }
\la{fweweffwef2}
\ee
Operating on equation (\ref{fweweffwef1}) with the derivatives $\pa^{\a \dot \e} 
\pa^{\z \dot \g}$ yields:
\be
0
=
-\pa^{\a \dot \e} 
\pa^{\z \dot \g}\pa_{\a \dot \b} 
\pa_{\d \dot \g}
W^{  \d \dot \b}_{{L}   
 p  }
-
    m^2 \ov d_{ p  q}  
\pa^{\a \dot \e} 
\pa^{\z \dot \g} 
 \ov W^{  q}_{{R} \a \dot \g }
\la{fweweffwef3}
\ee
 Then equation (\ref{fweweffwef2})  yields:
\be
0=
\D^2 W^{  \z \dot \e}_{{L}   
 p  }
-
    m^4 \ov d_{ p  q}  
 d^{ s  q}  
 W^{\z \dot \e}_{{L}  s   }\la{fweweffwef4}
\ee
Since the matrix $D_{p}^{\;\;s}=
\ov d_{ p  q}  
 d^{ s  q}  
$ is a positive definite hermitian matrix, it can be diagonalized, and this yields three equations of the form
\be
\lt ( \D^2 - m^4 D \rt ) 
\ov \w^{\a    }_{{L} }=0
\ee
for the three flavours, with $D$ different for each flavour.

These can be written
\be
\lt ( \D - m^2 \sqrt{D} \rt ) 
\lt ( \D + m^2 \sqrt{D} \rt ) 
\ov \w^{\a    }_{{L} }=0
\la{qerwfgegqerger1}
\ee
Only the term $\lt ( \D + m^2 \sqrt{D} \rt ) 
$ behaves like a normal equation of motion, giving rise to a normal propagator.  It has a zero at $m^2 \sqrt{D} $  for a real positive value of the energy squared $p_0^2 = m^2 \sqrt{D} + p_i^2$.  The other term $\lt ( \D - m^2 \sqrt{D} \rt ) 
$ has no zero for any real energy squared.  This unusual equation of motion is at the heart of the supersymmetry breaking mechanism in cybersusy\footnote{  For the baryons, this pattern repeats, except that there are  more factors like $\lt ( \D - m^2 \sqrt{D}  \rt ) 
$.}.

\section{ Construction of Compensators to build up an action invariant under the total BRS operator   $\d = \d_{\rm SS} + 
 \d_{\rm GSB} $ for the general case when the VEV is not zero }

Next  we will write down the action for the general case where 
the  term $\d_{GSB}\neq 0$. The operator $\d_{GSB} $ is defined in item \ref{deltaGSB} in section
 \ref{cyberalgebrasection}.

 Note that the scalar action ${\cal A}_{{\rm Scalar} }$, which consists of (\ref{ActionScalarL}), 
 (\ref{ActionScalarR}) and (\ref{ActionScalarMass}), is invariant under  $ \d_{\rm GSB} $ because $ \d_{\rm GSB} $  does not affect any of the scalar superfields.

The terms that are not invariant 
under $\d_{GSB}$ are the three parts of ${\cal A}_{{\rm   Dotspinor} }$, which are
 (\ref{ActionSpinorL}), (\ref{ActionSpinorR}) and (\ref{ActionSpinorMass}).

\subsection{ Construction of Left Compensators  ${\cal A}_{\rm  KCL1}$ and ${\cal A}_{\rm  KCL2}$}

We can compensate the noninvariance of the term 
${\cal A}_{\rm  DotspinorL}$
in equation (\ref{ActionSpinorL}) by adding two more terms ${\cal A}_{\rm  KCL1}$ and ${\cal A}_{\rm  KCL2}$ to the action. 
The subscript `KC' stands for `Kinetic   Compensator'. These three terms then form an invariant under the operator $\d_{L} = 
\d_{\rm SSL} + 
 \d_{\rm GSBL} $ in the following way:

\be
\d_{\rm SSL}  {\cal A}_{\rm  DotspinorL}
=0
\ee
\be
 \d_{\rm GSBL} {\cal A}_{\rm  DotspinorL}
+ \d_{\rm SSL} {\cal A}_{\rm  KCL1}
=0
\ee
\be
 \d_{\rm GSBL} {\cal A}_{\rm  KCL1}
+ \d_{\rm SSL} {\cal A}_{\rm  KCL2}
=0
\ee
\be
 \d_{\rm GSBL } {\cal A}_{\rm  KCL2}
=0
\ee

The superspace forms of ${\rm Action}_{\rm KCL1}$ and ${\rm Action}_{\rm KCL2}$ are non-local in superspace, and not very nice to look at.  However these expressions are local when they are written in components in spacetime:

\[
{\cal A}_{\rm KC{L}1}
=\int d^4 x \;
\lt ( 
{\ov p}^{q  p }
\ov \y^{}_{{L}     \dot \a q}
\D 
\w^{  \dot \a}_{{L}   
  p }
+
{\ov p}^{q  p }
\ov F^{}_{{L} q  }
\pa_{\a \dot \a}
W^{  \a \dot \a}_{{L}   
 p  }
\rt.\]
\be
\lt.
+ 
 {p}_{q  p }
 \y^{q  }_{{L}    \a}
\D 
\ov \w^{   p  \a}_{{L}   
 }
+
 {p}_{q  p }
 F^{ q }_{{L} }
\pa_{\a \dot \a} 
\ov W^{  p  \a \dot \a}_{{L}   
 }
\rt )
\la{formKCL1}
\ee

\be
{\cal A}_{\rm KC{L}2}=
\int d^4 x \;
 {p}_{q  p }
{\ov p}^{s  p }
\lt (
  A^{  q  }_{{L} }
\D 
\A^{ }_{{L}   s }
+
 F^{ q   }_{{L} }
 \oF^{ }_{{L}  s  }
\rt )
\la{formKCL2}
\ee
The result is that the total left action 
\be
{\cal A}_{\rm L\; Total}
= {\cal A}_{\rm  ScalarL}
+ {\cal A}_{\rm  DotspinorL}
+ {\cal A}_{\rm  KCL1}
+ {\cal A}_{\rm  KCL2}
\ee
is invariant under the left operator $\d_{L} = 
\d_{\rm SSL} + 
 \d_{\rm GSBL} $:
\be
\d_{L} {\cal A}_{\rm L\; Total}
=0
\ee
Of course, it is also invariant under the operator $\d_{R} = 
\d_{\rm SSR} + 
 \d_{\rm GSBR} $, but that is trivial because it is a function of the left fields only. 

Note that the signs of the terms  $A^{  q  }_{{L} }
\D 
\A^{ }_{{L}   s }$ and $
 F^{ q   }_{{L} }
 \oF^{ }_{{L}  s  }$
terms are the same in equation 
(\ref{formKCL2}), whereas they are opposite in 
the action (\ref{ActionScalarL}).    Note that 
${p}_{q  p }
{\ov p}^{s  p }$ is a positive hermitian matrix, which can be diagonalized to a positive diagonal matrix $P$. Then the total action, which is the sum  of (\ref{formKCL2}) and (\ref{ActionScalarL}), will have the right signs for these terms only if  $0\leq P< U $ where $U$ is the unit matrix. This issue is discussed more fully in \ci{cybersusyIV}.

\subsection{ Construction of Right Compensators  ${\cal A}_{\rm  KCR1}$ and ${\cal A}_{\rm  KCR2}$}

All the above discussion for the left sector is identical for the right sector, except for the change $L\ra R$ everywhere.
Here are the new parts for the right sector:

\[
{\cal A}_{\rm KC{R}1}=
\int d^4 x \;
\lt ( 
{\ov p}^{ {{p}}\dot q}
\ov \y^{}_{{R}     \dot \a { q}}
\D 
\w^{  \dot \a}_{{R}   
 {{p}}}
+
{\ov p}^{ {{p}}{ q}}
\ov F^{}_{{R} { q}  }
\pa_{\a \dot \a}
W^{  \a \dot \a}_{{R}   
{{p}} }
\rt.
\]
\be
\lt.
+ 
 {p}_{ {{p}}{ q}}
 \y^{{ q}  }_{{R}    \a}
\D 
\ov \w^{  {{p}} \a}_{{R}   
 }
+
 {p}_{ {{p}}{ q}}
 F^{ { q} }_{{R} }
\pa_{\a \dot \a} 
\ov W^{ {{p}} \a \dot \a}_{{R}   
 }
\rt )
\la{formKCR1}
\ee

\be
{\cal A}_{\rm KC{R}2}=
\int d^4 x \;
 {p}_{ {{p}}{ q}}
{\ov p}^{{{p}}{s } }
\lt (
  A^{  { q}  }_{{R} }
\D 
\A^{ }_{{R}   {s } }
+
 F^{ { q}   }_{{R} }
 \oF^{ }_{{R}  {s }  }
\rt )
\la{formKCR2}
\ee

\subsection{ Mass  Compensators  ${\cal A}_{\rm  MC1}$ and ${\cal A}_{\rm  MC2}$ do not exist}

This leaves just the dotspinor mass term 
(\ref{ActionSpinorMass}) to be compensated in order to achieve total invariance under 
\be
\d = \d_{\rm SS} + 
 \d_{\rm GSB}. 
\ee 
Naturally to complete the action, we want to construct       dotspinor mass compensator terms
 ${\cal A}_{\rm MC1}$  and  ${\cal A}_{\rm MC2}$
such that: 
\be
\d_{\rm SS} {\cal A}_{\rm Dotspinor\;  Mass}
=0
\ee
\be
 \d_{\rm GSB} {\cal A}_{\rm Dotspinor\;  Mass}
+
 \d_{\rm SS}  {\cal A}_{\rm MC1}
=0
\ee
\be
 \d_{\rm GSB}{\cal A}_{\rm MC1}
+
 \d_{\rm SS} {\cal A}_{\rm MC2} =0
\ee
\be
 \d_{\rm GSB}{\cal A}_{\rm MC2}  =0
\ee
However, it is easy to establish that no such local terms ${\cal A}_{\rm MC1} $ and ${\cal A}_{\rm MC2} $  exist 
\ci{cybersusyIV}.

\section{The Supersymmetry Breaking Action of Cybersusy}
\la{actionofcybersusy}
So we will now look at the following action 
\[
{\cal A}_{\rm Cybersusy} = 
{\cal A}_{\rm Susy} 
\]
\be
+ {\cal A}_{\rm  KCL1}
+ {\cal A}_{\rm  KCL2}
+ {\cal A}_{\rm  KCR1}
+ {\cal A}_{\rm  KCR2}
\la{cybersusyaction}
\ee
Putting all these together, we see that  ${\rm Action}_{\rm Cybersusy} $ consists of the terms
(\ref{ActionScalarL}), (\ref{ActionScalarR}), (\ref{ActionScalarMass}),  (\ref{ActionSpinorL}),
 (\ref{ActionSpinorR}), (\ref{ActionSpinorMass}), (\ref{formKCL1}), (\ref{formKCL2}), (\ref{formKCR1}) and  (\ref{formKCR2}).

All the terms except (\ref{ActionSpinorMass})
are invariant under the action of (\ref{qggrqegqerg1}):
More precisely we have: 
\[
\d {\cal A}_{\rm Cybersusy}=
\lt ( \d_{\rm SS} + 
 \d_{\rm GSB} \rt ) {\cal A}_{\rm Cybersusy}
\]
\be=
 \d_{\rm GSB} {\cal A}_{\rm Dotspinor\;  Mass}
\neq 0
\la{trouble}
\ee
From equation (\ref{trouble}) it is evident that there are
 two ways to get an action which satisfies
\be
\d {\cal A}_{\rm Cybersusy}=0
\la{qfewfwefwefwe}
\ee
These two ways are:
\ben
\item
Arrange that $\d_{\rm GSB}=0$, which means that gauge symmetry breaking is not present. For the theory with $
 \d_{\rm GSB} =0$, the kinetic compensator terms are absent and the most general action contains the terms 
${\cal A}_{\rm Dotspinor\;  Mass}$.  It also turns out that in this case, the action results in a supersymmetric spectrum and the dotspinor is now a massive supermultiplet.

\item
Arrange that the coefficient of
 ${\cal A}_{\rm Dotspinor\;Mass}$ is zero.
For the theory with ${\cal A}_{\rm Dotspinor\;  Mass}$ absent, the most general action contains kinetic compensator terms and it is invariant under the full operator $\d= \d_{\rm SS} + 
 \d_{\rm GSB} $.  It turns out that the action in this case also results in a supersymmetric spectrum and the dotspinor gives rise to a massless supermultiplet.
\een
We have been looking for an action for supersymmetry breaking, so let us examine the situation where we refuse to make this choice, and just keep the action ${\cal A}_{\rm Cybersusy}$.
We  note that ${\cal A}_{\rm Cybersusy}$ is uniquely chosen as the action which has the properties that
\ben
 \item
Supersymmetry breaking vanishes when the Vacuum Expectation Value (VEV) vanishes, and 
 \item
Supersymmetry breaking vanishes when the dotspinor superfield mass coefficent vanishes. 
\een

So we have arrived at a uniquely defined action for this symmetry, and it contains explicit supersymmetry breaking. The next questions are
\ben
\item
What is the spectrum?
\item
Does the theory make sense?
\een

\section{The polynomials that determine the masses}
\la{polysection}

Now let us see what the cybersusy action in section 
(\ref{actionofcybersusy}) yields for the masses. This is calculated in detail in \ci{cybersusyIV}.
The answer is surprisingly simple, given the complexity of the mixing.
The kinetic/mass terms for all the bosons and fermions in the above action can be inverted to find the propagators. Since they are quite mixed up, this is quite a chore. Our working hypothesis is that the mass eigenstates are the mass values of the momentum squared $p_{\m} p^{\m}$ which give rise to poles in those propagators. We define the dimensionless parameter:
\be
X = \fr{\D}{m^2}\Ra \fr{p_{\m} p^{\m}}{m^2}
\ee
and we combine the matrices above into the positive definite hermitian matrices:
\be
P = p p^{\dag};
G = g g^{\dag};
D = d d^{\dag}
\ee

Then it turns out that, for the simplest case of one flavour,   there are only three different kinds of poles in the Fermion propagator, and  the relevant masses arise from  the three negative real solutions of the following quintic polynomial equation for the spin $J= \fr{1}{2}$ fermionic leptons:

\be P_{\rm Quintic \; Fermi}(X) =  X 
 \lt \{ X^2 (  1-P) -  D \rt \}^2 
+      G  \lt \{ X^2    -  D \rt \}^2 =0
\la{wterjrtfuj2}
\ee

It also turns out that, for the simplest case of one flavour,  there are only four different kinds of poles for the Bose propagator, and  the relevant masses arise from the three  negative real solutions of the following quartic polynomial equation for the spin $J= 0$ bosonic scalar leptons:
\be
P_{\rm Quartic \; Bose }(X) = X^2  
\lt (  X  (1-P)^2  +  G  
\rt )^2 - \lt ( X(1-P^2)+G \rt )^2  D 
=0
\la{wterjrtfuj1}
\ee
and  the one negative real solution of the following quadratic polynomial equation for the spin  $J= 1$   vector boson lepton
\be
P_{\rm Quadratic \; Bose }(X) = X^2  - D=0 
\la{wterjrtfuj3}
\ee
This last equation is simply another way of writing the equation
(\ref{qerwfgegqerger1}) for the vector boson lepton.  The vector boson mass is not affected by the supersymmetry breaking mechanism of cybersusy.

Let us examine what happens to  the above polynomials 
(\ref{wterjrtfuj2}), (\ref{wterjrtfuj1})  and 
(\ref{wterjrtfuj3}) when we set the VEV to zero. 
Taking the VEV to zero is equivalent to setting $P=0$, and that results in:
\be P_{\rm Quintic \; Fermi}(X) \stackrel{P\ra 0}{\lra}
 \lt \{ X+G \rt \}  \lt \{ X^2    -  D \rt \}^2 
\ee
\be
P_{\rm Quartic \; Bose }(X) \stackrel{P\ra 0}{\lra}  \lt \{ X+G \rt \}^2  \lt \{ X^2    -  D \rt \} 
\ee
\be
P_{\rm Quadratic \; Bose }(X) \stackrel{P\ra 0}{\lra} X^2  - D 
\ee
 This is just ordinary supersymmetry as in section 
\ref{qegregregergre} above.

Now let us examine what happens to  the above polynomials 
(\ref{wterjrtfuj2}), (\ref{wterjrtfuj1})  and 
(\ref{wterjrtfuj3}) when we set the mass of the dotspinor multiplet to zero.  Taking the mass of the dotspinor multiplet  to zero is equivalent to setting $D=0$, and that results in:
\be P_{\rm Quintic \; Fermi}(X) \stackrel{D\ra 0}{\lra}  
X^4 \lt (  X  (1-P)^2  +  G  
\rt ) 
\ee
\be
P_{\rm Quartic \; Bose }(X) \stackrel{D\ra 0}{\lra}
X^2  
\lt (  X  (1-P)^2  +  G  
\rt )^2 
\ee
\be
P_{\rm Quadratic \; Bose }(X) \stackrel{D\ra 0}{\lra} X^2  
\ee
As claimed, the supersymmetry is restored, and there is a zero mass multiplet.  Note that the masses do depend on the gauge symmetry breaking parameter $P$ however.  

These details also confirm the remarks made in the  two points after equation (\ref{qfewfwefwefwe}) above.

\section{Three Numerical Examples}
\la{qfggqegerujujkuy}

Let us look at what this means in terms of particles. We have a quintic equation to solve to get the fermion masses, and a quartic equation and a quadratic equation to solve to get the boson masses. 

We solve these numerically using Mathematica for three examples, and here are the results, which we will discuss below in subsection \ref{wgqerrtghghe}.

\subsection{Choice 1 for  $P,G,D$--an example that probably is inconsistent with known experimental results}

For example, one finds that  the following choice for  $P,G,D$,
\be
P= .99999999, G = 1, D = 10^{18}
\ee
yields the following values for the zeros of the fermionic quintic polynomial  equation  $P_{\rm Quintic \; Fermi}(X)=0$:
\be
X=  -9.00722*10^{15} , -9.93414*10^{11},  -1,   
 \eb
   5.06707*10^{11}  - 8.65953*10^{11} i ,  
    5.06707*10^{11} + 
     8.65953*10^{11} i,
\ee
and the following values for the eigenvalues of the bosonic quartic polynomial  equation $P_{\rm Quartic \; Bose }(X) =0$:
\be
X=   -5.26316*10^7 , 
    1.9*10^{17}, -2.1*10^{17}, -4.7619*10^7
\ee

\subsection{Choice 2 for  $P,G,D$  }

Here is another choice for  $P,G,D$:
\be
  P= 1 - 10^{-15}, G = 1, D = 10^{30}
\ee
It yields the following values for the zeros of the fermionic quintic polynomial  equation:
\be
  -1.0000*10^{30},   -9.9999*10^{19},   -1.0000, 
\ee

\be
    5.0001*10^{19} - 8.6603*10^{19} i, 
    5.0001*10^{19} + 8.6603*10^{19} i, 
\ee
and the following values for the eigenvalues of the bosonic quartic polynomial  equation:
\be
  -1.0000*10^{15}, 
    1.0000*10^{30},  -3.0000*10^{30},   -3.3333*10^{14} 
\ee

\subsection{Choice 3 for  $P,G,D$  }

Here is another choice for  $P,G,D$:
\be
  P= 1 - 10^{-25}, G = 1, D = 10^{50}
\ee
It yields the following values for the zeros of the fermionic quintic polynomial  equation:
\be
  -1.0000\times 10^{50}, -2.1544\times 10^{33}, -1.0000,\eb
  1.0772\times 10^{33}-1.8658\times 10^{33} i, 1.0772\times
   10^{33}+1.8658\times 10^{33} i,\ee
and the following values for the eigenvalues of the bosonic quartic polynomial  equation:
\be
 -1.000\times 10^{25},1.0000\times 10^{50}, -3.0000\times 10^{50}, -3.3333\times
   10^{24} 
\ee

\subsection{Discussion of the  Choices for  $P,G,D$}
\la{wgqerrtghghe}

Suppose that we are  applying these choices to the electron in the simplest case where there is just one flavour.  Then:
\ben
\item
The model does make some sense in terms of the number of predicted masses, at least at first glance:
\ben
\item
For all three  choices, the fermionic quintic polynomial equation has three real negative solutions and two complex solutions. 
\item
For all three  choices, the bosonic quartic polynomial equation has    three real negative solutions and one real positive solution.
\item
For all three  choices, the bosonic quadratic polynomial equation has   one real negative solution and one real positive solution.
\item
The complex and positive solutions do not give rise to masses, for reasons explained after equation 
(\ref{qerwfgegqerger1}).  So these yield three fermionic masses and four bosonic masses.
\een
\item
The electron is the lightest particle in the broken supersymmetry multiplet:
\ben
\item
For all three  choices, the electron  ${\rm mass}^2$  would be the term corresponding to the fermionic solution $-1$, and the corresponding  ${\rm mass}^2$  would be $m^2$.  
\item
The electron has the lowest ${\rm mass}^2=m^2$ for all three  choices.
\item
It is encouraging, at least, that the electron is the lowest mass particle for these three choices of the parameters $P,G,D$, and that the other particles are very much more massive.  
\een
\item For all three choices,  
the next lightest particle in the broken supersymmetry multiplet is a scalar selectron, and its mass is very different for the three choices:
\ben
\item

For Choice 1, the next lowest  ${\rm mass}^2$, after  the electron, would be the lowest mass bosonic lepton, which would have  ${\rm mass}^2$ $=4.7619*10^7 m^2$, corresponding to ${\rm mass}$ $ =  6900.66 m$. So the mass of this lowest mass leptonic boson  would be predicted to have a mass a little larger than the mass of three protons, since the proton has a mass of about 2000 times the mass of the electron.  No doubt, such a particle would have been found by now if it existed. 
\item
For Choice 2, the next lowest  ${\rm mass}^2$, after  the electron,   would be the lowest mass bosonic lepton, which would have  ${\rm mass}^2$  $= 3.3333*10^{14}  m^2$, corresponding to ${\rm mass}$  $ =  1.82574*10^7 m$. So, for Choice 2, the mass of this lowest mass leptonic boson  would be predicted to have a mass of about  the mass of 10,000 protons, since the proton has a mass of about 2000 times the mass of the electron.  \item
For Choice 3, the next lowest  ${\rm mass}^2$, after  the electron,   would be the lowest mass bosonic lepton, which would have  ${\rm mass}^2$  $= 3.3333*10^{24}  m^2$, corresponding to ${\rm mass}$  $ =  1.82574*10^{12} m$. So, for Choice 2, the mass of this lowest mass leptonic boson  would be predicted to have a mass of about  the mass of $10^9$ protons, since the proton has a mass of about 2000 times the mass of the electron.  
\item
  Evidently, there is a range between these masses for the lightest supersymmetric partner (3 to $10^9$ proton masses) that is amenable to detection at the LHC, so, if the numbers work out, one could detect supersymmetry at the LHC consistent with this model for supersymmetry breaking.
\een
\item
The other particles are all heavier, but they are also quite constricted by the masses of the electron and the lightest selectron, since there are only three parameters $P,G,D$ and we already have two masses:
\ben

\item
The other two negative  fermionic  solutions would correspond to heavy leptonic fermions with ${\rm mass}^2=9.00722*10^{15} m^2$ and  ${\rm mass}^2=9.93414*10^{11} m^2$ for Choice 1 and much heavier masses for Choice 2.

\item
For Choice 1, the other two bosonic masses here would be predicted to have ${\rm mass}^2=5.26316*10^7 m^2$ and
 ${\rm mass}^2=2.1*10^{17} m^2$.  Again these are much heavier for Choice 2. 
\item
For Choice 1, there would also be the observable leptonic vector boson with ${\rm mass}^2=\sqrt{D} m^2= 10^{9} m^2$ so the mass would be ${\rm mass}=\sqrt{\sqrt{D} m^2}= \sqrt{10^{9} m^2}= 31623 m$
\item
For Choice 2, there would also be the observable leptonic vector boson with ${\rm mass}^2=\sqrt{D} m^2= 10^{15} m^2$ so the mass would be ${\rm mass}=\sqrt{\sqrt{D} m^2}= \sqrt{10^{15} m^2}=3.1623*10^7 m$
\item
For Choice 3, all the other masses are very large.
\item
Presumably one can get the masses (other than the electron)  higher than this too. But the thorough exploration of the set of possible solutions does not look at all simple, and  I shall not attempt any phenomenology here.
\een
\item
The theory also appears to make sense in terms of counting degrees of freedom, at least at first glance:

\ben
\item
It appears that  a big  mass splitting between the electron and the other particles requires that the term $P$ be very near to one (note  that $0\leq P <1$ seems indicated by the form of the action) and that the closer it is, the greater the splitting is.  
\item
It appears as though the closer P is to one, the greater the gauge symmetry breaking is, in some sense.  So it seems that the greater supersymmetry breaking requires the greater gauge symmetry breaking.  It would be nice to understand this better.
\item
There are three $J= \fr{1}{2}$ fermions, three $J= 0$ scalar bosons, and one $J= 1$ vector boson.  The three fermions get mixed together to give three  new masses. The third scalar mass arises from  the longitudinal parts of the vector boson, and there is a mixing of the three scalars to give three different masses for the scalars.  The vector boson itself does not change mass when supersymmetry breaks or when gauge symmetry breaks.
Note that this is very different from the Higgs-Kibble-Guralnik mechanism in spontaneously broken gauge theories, though it does have some things in common too. 
\item
It is encouraging that seven masses arise from three parameters for one flavour,  and that there are not more masses than one could comfortably assign to particles.   If, for example, the fermion polynomial gave four masses instead of three, that would be a disturbing result.
\item
The above discussion is for one flavour, of course. For three flavours, if all the matrices $g,d,p$ and $\ov g, \ov d, \ov p$ are diagonal at the same time, one simply repeats the above pattern with different numbers for each flavour. 
In general however it is more  complicated.  The full result is given in \ci{cybersusyIV}, and that result needs  analysis
that looks rather challenging.  There are also angles, and probably phases, to take care of in that case.
 \een
\item
 It must be admitted that the necessary choice of parameters does not seem `natural' in the technical sense--it appears that some of the parameters must be chosen to be very large or very small numbers.  That is not surprising of course.  The electron and neutrino masses are very small, and no supersymmetric partners have yet been discovered.  So  the masses of the superpartners must be very large. On the other hand, it is nice that this is at least possible, because of the special form of the polynomials that determine the masses. Because the supersymmetry breaking has some features in common with explicit breaking, there is no reason to expect any mass sum rules \ci{weinberg3.192}.
\item
It is tempting to wonder whether the dotspinor mass comes from the superstring in some way.  

\een

\section{ An informal introduction to the development of this mechanism}
\la{history}

This paper had its origins in the effort by this author and some collaborators
to compute the cohomology of the chiral supersymmetry model of Wess and Zumino \ci{WZ} using spectral sequences \ci{general}.

  Early results using just the chiral transformations, but without including Zinn sources, 
showed that the ghost charge one space was  huge, but largely consisted of objects with unsaturated spinor indices \ci{holes}
\ci{dixprl} 
\ci{kyoto}
\ci{dixmin}
\ci{dixminram}. 
 Other authors concentrated on the cohomology without unsaturated spinor indices, because of course the action has no unsaturated spinor indices 
\ci{Rupp} \ci{Brandt}.

It was clear that if anything interesting was going to happen in the sector with unsaturated spinor indices, one had to calculate the BRS cohomology including the 
Zinn sources \ci{ZJ}.  
It turns out that there is plenty of cohomology at ghost charge zero when the Zinn sources are included, and it is closely related to the ghost charge one cohomology. This will be discussed in \ci{cybersusyII}.

 It is well known \ci{brs} that the gauge anomalies can be formulated in the following way
\be
\d_{BRS} {\rm Action}
= {\rm Anomaly}
\la{ewqrrewfwe}
\ee
where $\d_{BRS}$ is the nilpotent BRS operator for the theory and {\rm Action} is the usual integral of the Lagrangian, and  Anomaly is an  integrated local polynomial in the fields, with the same quantum numbers as the Action, except that it has ghost charge one. 

The natural question is whether the cohomology of supersymmetry gives rise to a comparable anomaly.  It turns out that the cohomology of the chiral Wess Zumino action at ghost charge zero contains an infinite set of composite chiral dotted spinor superfields ${\widehat \w}_{\dot \a}$.  There is also an infinite set of ghost charge one composite objects of the form     ${\widehat A} \; \ov C_{\dot \a}$ where   ${\widehat A}$ are composite chiral scalar superfields and $  \ov C_{\dot \a}$  is the constant supersymmetry Weyl spinor ghost.

By analogy with the way that the usual chiral anomalies work in (\ref{ewqrrewfwe}), a supersymmetry anomaly would be expected to have the form:
\be
\d_{BRS} {\widehat \w}_{\dot \a}=   {\widehat A} \; \ov C_{\dot \a}
\la{twetgghlkhre1}\ee 

There were two immediate problems here:
\ben
\item
 The first problem was to understand the BRS cohomology of the situation in detail and in full, including gauge theory, and that is very hard indeed. An effort to do this was made in  \ci{bigpaper} and the useful results from that will be used here and explained   in \ci{cybersusyII} and \ci{cybersusyIII}. There is a pressing need for a better paper on this topic, but that paper is not easy to write, and it is likely to be impossible to read, and there are many unsolved problems and gaps. However, although there is still much to learn, a large set of the possible 
${\widehat \w}_{\dot \a}$ and $  {\widehat A} \; \ov C_{\dot \a}$ are now known.
\item
The other problem was to do a calculation of an anomaly if one exists. It gradually became fairly clear to the author that supersymmetry anomalies like this did not seem to arise in one-loop calculations.
\een
  At the same time, the results indicated that the
composite chiral dotted spinor superfield operators of the form ${\widehat \w}_{\dot \a}$ that were in the cohomology space, were a very different kind of operator from the analogous gauge invariant operators in gauge theories.  In the supersymmetric theory, there is a constraint that must be satisfied to make ${\widehat \w}_{\dot \a}$, and the existence of these operators   has a fundamental dependence on the equations of motion.  There is nothing like ${\widehat \w}_{\dot \a}$ in gauge theories without supersymmetry.

It was then discovered that 
the standard supersymmetric model was a useful area to look at the composite chiral dotted spinor superfield operators of the form ${\widehat \w}_{\dot \a}$ that were in the cohomology space.  It turns out that the SSM has lots of examples of solutions of the constraint that ${\widehat \w}_{\dot \a}$ must satisfy.  Even more interesting is the fact that  this is related to the peculiar  structure of the SSM, with its direct products of groups and its left-right asymmetry 
\ci{cybersusyIV}.

Even more remarkably, those examples of ${\widehat \w}_{\dot \a}$ look like composite particles of a familiar kind. A first effort at making a fairly complete list of these  was made in  \ci{hadronpaper}, but those results are  incomplete, and the results in this paper and in \ci{cybersusyII} and \ci{cybersusyIII} are more complete and more accurate.
A more complete list of these examples needs to be made, including proton and neutron type hadrons to start with.

The first real progress towards a result that has some physical predictions in it, was made  after  \ci{hadronpaper} was written.    The author noticed that  there was actually something that looked just like a supersymmetry anomaly, except that it had powers of the mass parameter in front of it, and also, it did not come from a one-loop quantum calculation like the gauge anomalies in (\ref{ewqrrewfwe}).  Instead, it appeared at tree level when the gauge symmetry breaking was turned on.  It had the form 
\be
\d_{BRS} {\widehat \w}_{\dot \a}= m^n {\widehat A} \; \ov C_{\dot \a}
\la{twetgghlkhre}
\ee
where $m^n$ is some power of the mass parameter in the theory.

The question then was what to do with this result.  The only idea that came to the author was  to convert  the relevant composite fields to elementary effective fields while conserving the same algebra (\ref{twetgghlkhre}), and then write down a Lagrangian field theory for those fields with that invariance algebra.
This yields an effective action. That creation of effective fields, an effective algebra and an effective action, is the origin of cybersusy.

 The effective BRS operator actually contains (\ref{twetgghlkhre1}), but with elementary effective fields.  The mass parameters must be removed as a result of the canonical dimensions that one needs for the effective theory. 
  Because the equation 
(\ref{twetgghlkhre1}) really is effectively a supersymmetry anomaly, it is not surprising that the effective action has some peculiar features from the point of view of supersymmetry.

It is also interesting and convenient that the dotspinors that are generated as effective fields do actually have a rather interesting and unusual action.  Furthermore, that action fits well into the idea that these dotspinors represent a new kind of supersymmetry multiplet corresponding to some supersymmetric set of bound states.  These actions  have not been used in supersymmetry before, as far as the author knows.

But the real surprise is that this effective action   gives rise to another issue, namely--explicit supersymmetry breaking. If one starts with the effective action with a zero vacuum expectation value, and tries to turn the VEV on  while preserving supersymmetry in the effective action, one has to shut off the mass of the dotspinor. Any VEV at all is incompatible with any mass for the dotspinor, if one wants to preserve supersymmetry.  On the other hand, if the VEV is zero, then the dotspinor mass conserves supersymmetry.  It is impossible to have the following three things at the same time:
\ben
\item
 a non-zero mass for the dotspinor, 
\item
 a non-zero VEV that breaks gauge symmetry, and 
\item
unbroken supersymmetry.
\een
but one can have any two of them.

 This situation is a bit reminiscent of what happens with  gauge anomalies, but there are also lots of differences. 
Gauge anomalies also arise from an incompatibility of requirements that, naively at least, seem  compatible.

It appears that the theory behaves very nicely no matter which two of the above three possibilities are chosen.

In particular if one chooses to give supersymmetry up, by keeping a non-zero dotspinor mass, and keeping gauge symmetry breaking with a non-zero VEV, the theory still behaves nicely and supersymmetry breaking is generated in a unique way. 
The details and consequences of that have been explored in this paper. The result seems to be that a kind of minimal and unique explicit supersymmetry breaking is generated by this procedure, and it is linked strongly to the SSM.

\section{Conclusion}

\subsection{Summary}

This  paper gives a brief survey of the results that are explained at greater length in the sequels \ci{cybersusyII}
\ci{cybersusyIII}
\ci{cybersusyIV}.

We have seen that, using fairly reasonable assumptions,  cybersusy yields a unique effective action for leptonic supersymmetry breaking. Furthermore that action
  yields a spectrum for the leptons and their superpartners that is well defined, and which does not seem to get into any of the trouble that other supersymmetry breaking schemes encounter. The neutrino works identically, and the baryons look rather similar, although more complicated.  It remains to be seen whether similar methods can deal with the other aspects of supersymmetry breaking, particularly the baryons to start with.

\subsection{Some Remarks about the name cybersusy}
\la{fsdfsdfdfsdf}

The name cybersusy is meant to refer to the complicated structure that underlies this method for understanding supersymmetry breaking.  It is composed of words referring to the various aspects involved, including such things as
BRS {\underline c}ohomolog{\underline y}, gauge and supersymmetry
 {\underline b}reaking, {\underline e}ffective actions, and 
also {\underline r}etroversion in the context of
 \underline{su}per\underline{sy}mmetry.  
The retroversion refers to the fact that the cohomology also generates a supersymmetric version of baryons, composed of quarks, which indicates that cybersusy is poised to 
give supersymmetric breaking in the baryonic mass spectrum, rather than that of quarks. 

 This is amusingly retroverted, because supersymmetry did arise from the hadrons, through Regge trajectories, dual models, the string and the superstring \ci{history}. It is not yet known whether  cybersusy gives reasonable results for the baryons, but finding out is a straightforward  (but laborious) extension of the present results.    Moreover there is a reasonable chance that it works for baryons, because parts of that algebra are the same as the algebra here, and the observed baryons occupy  positions in the baryonic algebra and effective action that are similar to the position of  the electrons  in the present paper. 

Cybersusy arises from the BRS cohomology of chiral supersymmetry as applied to the SSM.  The relevant parts of that cohomology have  been found using spectral sequences in Fock Space.  It is probably significant that the solution to the cohomology has been found using component fields, with the Zinn sources for the variations of the components, and with the auxiliaries integrated \ci{cybersusyII}.  This removes manifest supersymmetry.  It appears to the author that this  cohomology, which gives rise to the composite dotspinors, cannot be easily or locally expressed using superspace, and that it is probably very hard to recover the same results using superspace. 

 The properties of the dotspinor superfields, and their strange quartic  equations of motion, and the flexible formulation of the philosophy of effective actions in section \ref{fqwfwefwfwef}, are the origin of the supersymmetry breaking in cybersusy.  The dotspinors  are also ubiquitous in the BRS cohomology of the chiral superfield \ci{cybersusyII}.  The SSM allows solutions of the BRS cohomology constraint equations for dotspinors, and these solutions are closely related to the particles we see \ci{cybersusyIII}.   It is also remarkable that the equations 
(\ref{ewrwerwerP}) and (\ref{ewrwerwerE}), which give rise to
(\ref{deltaGSBL}) and (\ref{deltaGSBR}), are exactly what one would expect from the cohomology for an anomaly in the composite theory, except for the mass factors and the fact that no loop computation is needed.  

The composite dotspinors that emerge from the cohomology do transform like superfields, but they are certainly not composed of superfields \ci{cybersusyII}.  The equations of motion, which are implicit in the Zinn sources which are included in the composite dotspinors, play an essential role in the composite dotspinors and the constraints that they satisfy, and also in the new algebra that they generate after gauge symmetry  breaking \ci{cybersusyIII}. 

 It is also noteworthy that although the new cybersusy algebra used in this paper to break supersymmetry can be expressed in terms of superfields,   the action for the lepton mixing in this paper  is non-local when expressed in superspace, and it also contains explicit supersymmetry breaking factors of $\q$ and $\ov \q$.  It seems simplest to deal with it using components fields.

It should be borne in mind that  any correct result that is found using components must be preferred, even if it conflicts with some superspace prediction or understanding, because superspace is only a way of making the theory more compact. If there is a conflict, we are forced to prefer the component formulation of supersymmetry. We seem to have come to a point where components are playing an important role again.  

It should also be remembered that gauge anomalies can destroy unitarity in gauge theories, where the gauge invariance is a local symmetry.  The `effective supersymmetry anomalies' that arise in the cybersusy algebra do indeed destroy 
supersymmetry, in their own special way, but since supersymmetry is not a local symmetry (as far as the standard supersymmetric model is concerned, at least), this does not cause any problem with unitarity.

For  the superstring, it is also necessary to write down effective actions to try to understand what is going on when one considers particles.  It would be good to have some information flow between cybersusy and the superstring.  Does the superstring give rise to the dotspinor masses?

\subsection{Does supersymmetry break when the vacuum condenses?}

We have seen that the  dotspinor mass term   results in supersymmetry breaking, and that this breaking arises as soon as the VEV which breaks gauge symmetry becomes non-zero. In superspace notation, this dotspinor mass term, is, for one flavour: 
\be
\sqrt{D} \; m^2 \int d^4 x \; d^2 \q \; {\hat \w}^{\dot \a }_{\;{\rm L }} {\hat \w}_{\dot \a \;{\rm R }}
\ee
 We have also seen that there are choices of the parameters $R,D,G$ such that the supersymmetry breaking is huge and such that the mass splitting between the electron (or neutrino), and the lowest mass superpartner, is probably as large as we want.  Moreover it seems to be natural to have the electron (or neutrino) as the lowest mass particle.

It is somewhat plausible that the early evolution of the universe actually took the course of this mechanism. 

One could imagine that at very high temperatures, at some extremely short time after the Big Bang, before the condensation of the vacuum through the emergence of the 
VEV \ci{inflationstuff}, the universe was described by the supersymmetric standard model, possibly together with other dotspinor masses from the superstring.  That might imply infrared slavery, and so would imply the existence of  bound states. The cohomology of composite operators appears to be telling us  the structure of the multiplets which could be expected to play the role of  the bound states.  The cohomology also tells us how these bound state multiplets transform after gauge symmetry breaking occurs through condensation. The  cohomology for the baryons shows similar features.

\subsection{Some Puzzles}

Here are a few of the many features of this construction that puzzle the author:
\ben
\item
From the way this supersymmetry breaking works, it appears that the vacuum remains at a zero energy level even after supersymmetry breaking.  Is this correct?
\item
What can one say about heavy leptons, heavy neutral leptons, dark matter etc. in view of this mechanism?
\item
What would be an efficient way to look for evidence of this mechanism? 
\item
What happens for the hadrons, all the baryons and mesons? Evidently it will be interesting to see if the model makes any sense when applied to these. 
\item
What happens for the other particles, such as the Higgs and the weak vector bosons and the photon? 
\item
Can one extend the new action to an interacting action and calculate radiative corrections?  Or does one use the other dotspinors in the cohomology for that? 
\item
What does this say about the superstring?

\item
Does the superstring set the scale for
the mass parameters $G$ and $D$ and $R$? How?

\item
If this mechanism is correct, does that mean that supergravity is not broken?  Could there be massless gravitinos?
\item
Do we get some useful information about bound states  from this construction?  For example does it tell us something useful about baryons? 
\item
Does cybersusy fit well with inflation scenarios? 
\item
How tightly is cybersusy bound to the SSM? Can it be extended to other models? 

\een

\subsection{Theoretical Work to be Done}

It would be interesting to study the mass spectrum of the hadrons, including the mesons, and to complete the cohomology with gauge theory included to see whether the vector bosons $W^+,W^-, Z$ and $A$, for example, behave in a proper way upon supersymmetry breaking using the present ideas.

Just using the present results about the cohomology of chiral supersymmetry, one can test to see whether
the superpartners of the baryons are heavier than the observed baryons,  which is necessary for the theory to agree with experiment.  

It is fairly clear that there may be enough cohomology in the gauge theory (it works rather like the present case) to generate  a similar mechanism for the vector bosons. Assuming cybersusy exists for the gauge theory, then the tests are as follows: 
\ben
 \item
The gauginos and other superpartners of the vector bosons need to be heavier than the gauge particles.  This is necessary for the theory to agree with experiment.  
\item
The superpartners of the hadronic mesons need to be heavier than the observed mesons.  This is also necessary for the theory to agree with experiment.  
\item
For  the Higgs mesons, it is not clear 
what  is necessary for the theory to agree with experiment, since they have not yet been seen.  It is conceivable that the superpartners of the scalar Higgs are lighter than the Higgs, as far as things stand at present. 
\een

From the general BRS cohomology of chiral supersymmetry, we also know that the dotspinor is certainly not the only relevant leptonic composite operator in the cohomology space for a given set of quantum numbers.  What about the others?  It does appear that any other relevant operators have more derivatives in them, and are therefore suppressed by powers of $\fr{\rm momentum}{\rm mass}$, where the mass is likely to be very large.
More work is needed on the cohomology.

\subsection{Synopsis of the next papers}

\la{gegoprgejio}
  Cybersusy  is based on the BRS cohomology of composite operators in the   Supersymmetric Standard Model (the SSM) with Gauge Symmetry Breaking (GSB).

First, in \ci{cybersusyII}, we give an introduction to the BRS cohomology of the massless Wess Zumino chiral action for a superscalar multiplet.  Our main concentration here is on the existence of what we call simple composite dotspinor multiplets and their constraint equations.  These  multiplets can only be discovered using the cohomology, and are hidden from view if one uses superspace only.

Then, in \ci{cybersusyIII}, we solve the constraint equations discussed in  \ci{cybersusyII},   using the massless supersymmetric standard model for the example. We note that the solutions are familiar for the hadronic sector--they closely resemble a supersymmetric quark model for  the hadrons.  For the leptonic sector, something analogous, but new, takes place.

Also in \ci{cybersusyIII}, we introduce the term which gives rise to the Vacuum Expectation Value (VEV) which generates gauge symmetry breaking in the SSM,
and note the new algebra generated by it among the leptonic composite dotspinor operators found in \ci{cybersusyIII} (the leptons are simpler to analyze than the hadrons). The VEV breaks the gauge symmetry from $SU(3) \times SU(2) \times U(1)$ to $SU(3)\times U(1)$.  As is well known in the SSM,    before and after gauge symmetry breaking, the theory exhibits unbroken supersymmetry for the elementary fields which appear in the action.  And as long as supersymmetry is not spontaneously broken, the vacuum energy remains zero \ci{ferrara}.

Then in \ci{cybersusyIV} we examine the effective action which arises from the new algebra for the composite dotspinor fields and note that because of the dotspinor nature of some of the new effective fields, there is an induced explicit supersymmetry breaking in the effective action as soon as the gauge symmetry breaking occurs. The polynomials
referred to in section  \ref{polysection} of this first paper are derived from the cybersusy action in \ci{cybersusyIV}.

\subsection{Implications for Experiment}

At present, the author does not see how to efficiently go about devising a test for these theoretical predictions, or how to distinguish cybersusy experimentally from other methods for breaking supersymmetry.   These are very important questions, of course, particularly given the advent of the LHC at CERN.

Of course, if experiments at the LHC  could actually  find the 9 spin $J=\fr{1}{2}$ leptonic fermions, and the 9 spin $J=\fr{1}{2}$ leptonic scalar bosons, and the 3 spin $J=1$ leptonic vector bosons, as predicted by cybersusy for the electron with three flavours, then there would be significant correlations to be observed.  We know, that for one flavour, cybersusy predicts seven different masses from the three real positive parameters $G,D,R$, and a comparable statement must also be true for three flavours.

\vspace{.2in}

\hspace{1.6in} {\bf Acknowledgments}
\vspace{.2in}

I thank J.C. Taylor and Raymond Stora for stimulating correspondence and conversations over many years.

\end{document}